\documentclass[twocolumn]{aastex63}

\usepackage{booktabs}
\usepackage{xspace}

\newcommand{\pow}[1]{\ifmmode{}^{#1}\else ${}^{#1}$\fi}

\newcommand{\cm}{\,\ifmmode{{\mathrm{cm}}}\else cm\fi}

\newcommand{\ergps}{\,{\rm erg}\,{\rm s}\ifmmode{}^{-1}\else${}^{-1}$\fi}
\newcommand{\Mpch}{\,{\rm Mpc}\,\ifmmode{} h^{-1}\else $h^{-1}$\fi}
\newcommand{\snru}{\,\ifmmode{\mathrm{Myr}^{-1}}\else Myr${}^{-1}$\fi}
\newcommand{\kms}{\,\ifmmode{\mathrm{km}\,\mathrm{s}^{-1}}\else km\,s${}^{-1}$\fi\xspace}

\newcommand{\lya}{Ly$\alpha$\xspace}
\newcommand{\Lya}{Ly$\alpha$\xspace}

\newcommand{\halpha}{\mbox{H$\alpha$}}
\newcommand{\hbeta}{\mbox{H$\beta$}}

\newcommand{\hi}{\mbox{H{\sc i}}}

\newcommand{\msunyr}{\mbox{M$_\odot$~yr$^{-1}$}}

\newcommand{\zsys}{\mbox{$z_\mathrm{sys}$}}

\shorttitle{The LASD}
\shortauthors{Runnholm et al.}
\submitjournal{PASP}

\begin{document}
\title{The Lyman Alpha Spectral Database (LASD)}
\correspondingauthor{Axel Runnholm}
\email{axel.runnholm@astro.su.se}

\author[0000-0002-1025-7569]{Axel Runnholm}
\affiliation{Department of Astronomy, Oscar Klein Centre, Stockholm University, AlbaNova universitetscentrum, SE-106 91 Stockholm, Sweden}

\author[0000-0003-2491-060X]{Max Gronke}
\altaffiliation{Hubble Fellow}
\affiliation{Department of Physics \& Astronomy, Johns Hopkins University, Baltimore, MD 21218, USA}
\affiliation{Department of Physics and Astronomy, University of California, Santa Barbara, 93106, USA}

\author{Matthew Hayes}
\affiliation{Department of Astronomy, Oscar Klein Centre, Stockholm University, AlbaNova universitetscentrum, SE-106 91 Stockholm, Sweden}

\begin{abstract}
 Lyman $\alpha$ (Ly$\alpha$) emission from star-forming galaxies is an important tool to study a large range of astrophysical questions:  it has the potential to carry information about the source galaxy, its nearby circumgalactic medium, and also the surrounding intergalactic medium. Substantial observational and theoretical work has therefore focused on understanding the details of this emission line. These efforts have been hampered, however, by an absence of spectroscopic reference samples that can be used both as comparisons for observational studies and as critical tests for theoretical work. For this reason, we have compiled a large sample of Ly$\alpha$ spectra, at both low and high redshift, and created a publicly available online database, at \href{http://lasd.lyman-alpha.com}{\texttt{lasd.lyman-alpha.com}}. The \emph{Lyman Alpha Spectral Database} (LASD) hosts these spectra, as well as a large set of spectral and kinematic quantities that have been homogeneously measured for the entire sample. As part of this we have developed an automated redshift determination algorithm which we show is accurate to within less than $\pm$180kms$^{-1}$ on average, across many different Ly$\alpha$ profiles. The measurements can conveniently be viewed online and downloaded in tabular form. The LASD has the capacity for users to easily upload their own Ly$\alpha$ spectra, and all the same spectral measurements will be made, reported, and ingested into the database. We actively invite the community to do so, and the LASD is intended to be a long-term community resource. In this paper we present the design of the database as well as descriptions of the underlying algorithms and the initial Ly$\alpha$ emitter samples that are in the database. 
\end{abstract}


\section{Introduction}\label{sec:intro}

The Lyman alpha (\lya) emission at 1215.67\,\AA\ originates from the $n=2-1$ transition of atomic hydrogen, where $n$ is the principal quantum number. \lya\ is intrinsically the strongest spectral line of astrophysical nebulae. The line strength combined with the  restframe UV wavelength means that it becomes a readily observed beacon from high redshift sources. Indeed \lya\ has seen extensive, and very successful, use for detection of high redshift galaxies in both narrowband \citep[e.g.,][]{2000ApJ...545L..85R,2008ApJ...681..856R,2014ApJ...797...16K} and spectroscopic \citep{2004ApJ...606..683S, 2005MNRAS.359..895V, 2017MNRAS.471..267D, 2007ApJ...663...10S, Herenz.2017mw, Urrutia.2019} surveys. 
Additionally, at these high redshifts, the \lya transition is often the only observable spectral line in the observer-frame optical and is therefore commonly used for spectroscopic confirmation of very high redshift galaxies detected by dropout techniques. 
Furthermore, since the intergalactic medium (IGM) becomes more neutral and, thus, more opaque to \lya\ photons towards higher redshifts, the (non)detection of \lya\ emitting galaxies provides us with tight constraints on the progress of reionization \citep{2014PASA...31...40D,Mason2019}.

Apart from a pure detection tool, the power of \lya\ lies in its resonant nature and consequently in its susceptibility to neutral hydrogen \textit{within} the emitting galaxy or in close proximity shaping the emergent \lya\ observables.
However, interpreting \lya\ emission from galaxies in terms of physical properties of the system or even using it for precise redshift determination is not trivial. This is because the \lya\ transition is resonant and therefore \lya\ photons experience extensive scattering in, and interactions with, the surrounding medium when escaping from virtually any environment \citep{1973MNRAS.162...43H,1990ApJ...350..216N}. This means that the emergent spectral line profile carries with it an imprint of the medium through which it travels, making it very complex but also potentially very informative of the physical conditions in the galaxy.

For this reason there has been extensive work done, both empirical \citep[e.g.,][see \citealt{2015PASA...32...27H} for a review]{2017A&A...597A..13V,Rivera-Thorsen.2015,2020ApJ...892...48R} and theoretical \citep[e.g.,][]{1990ApJ...350..216N,2001ApJ...554..604A,Verhamme2006,2017A&A...607A..71G, 2020MNRAS.497.3925L}, to attempt to decode or model \lya\ and determine what the imprint of various physical properties of the galaxy is on the line.  Community-wide, however, there are major difficulties in the interpretation of these results: individual observational samples of \lya\ emitting galaxies are often small, have been assembled in a piecemeal fashion, and different researchers have made different sets of measurements, using various definitions and methodologies/algorithms.  For empirical studies this means that recovered properties may not be comparable, correlations have small statistical significance, and robust conclusions are hard to draw. For theoretical studies on the other hand it means that empirical reference samples, against which the models can be tested, are hard to come by, which complicates sanity checks of model outputs. 

The main quantities derived from the \lya\ spectra reflect either photometric values (flux, luminosity, EW) or kinematic properties (e.g. velocity shifts).  Further quantities such as various asymmetry and skewness measures are weighted combinations of both wavelength and flux axes of in the one-dimensional spectra. For example simple velocity offsets of
the main (usually red) \lya\ peak are frequently derived. See for instance 
\citet{Steidel.2010,Hashimoto.2013,McLinden.2014,Rivera-Thorsen.2015}, all of
which are derived by ascribing a characteristic velocity to the \lya.  This
characteristic velocity may be derived from a Gaussian fit to the line, the
velocity of the peak emission, the first moment measured over a certain window,
or possibly other definitions.  As \lya\ is redistributed in velocity space by
scattering in galaxy winds and the IGM, asymmetry measurements have often been
employed.  The `class' of asymmetry measurements has over the years included
parametric fitting of split-Gaussian profiles, non-parametric measurements of flux distributions bluewards and rewards of line
centre (when $z_\mathrm{sys}$ is known, e.g. \citealt{Erb.2014}) or with
respect to the maximum flux (when $z_\mathrm{sys}$ is unknown, e.g.
\citealt{Rhoads.2003}), or recast estimates of the skewness statistic
\citep{Shimasaku.2006,Kashikawa.2006}.  All of these measurements differ
between application, group, and historical precedent and, moreover, will
further depend upon what rest-wavelength/velocity window is used for the
calculation, inclusion of errors, etc.  

In this work we present the `Lyman Alpha Spectral Database' (LASD), the  goal of which is to help resolve some of the issues described above. The database and its associated website \url{http://lasd.lyman-alpha.com} allow the community to upload calibrated \lya\ spectra which will be processed through a homogeneous analysis pipeline. In this paper, we present the database structure, web interface, and the analysis pipeline in Sec.~\ref{sec:Methods}. We describe the initial dataset consisting of  $\sim 340$ publicly available \lya spectra in Sec.~\ref{sec: Dataset}, and present some tests and some example correlations in Sec.~\ref{sec: Results}.  We present some concluding remarks and discuss the outlook for the LASD in Sec.~\ref{sec:conclusions}.


    \begin{figure*}
      \centering
      \includegraphics[width=\linewidth]{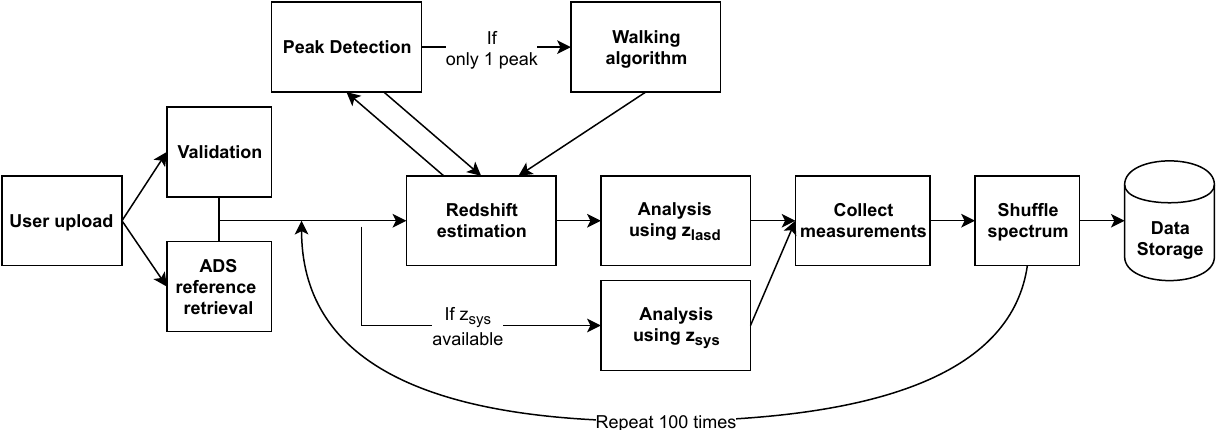}
      \caption{Visual representation of the LASD pipeline from user submission to final storage in the database.
      }
      \label{fig:lasd pipeline}
    \end{figure*}

\section{Methods}\label{sec:Methods}
\subsection{Database \& Web Interface}

The LASD is built entirely in python using the \texttt{Django}\footnote{\url{https://djangoproject.com}} web framework both to deliver the user interface and manage the \texttt{PostGreSQL} database.

The database is structured into the following three primary tables: 
\begin{enumerate}
    \item Observations: This holds all the raw data that was uploaded by the user as well as the unpacked and calibrated spectrum. Note that not all of this data is available to the user (see below) but storing the uploaded spectra allows us to reanalyze them in the future, e.g., to introduce new measurements. 
    \item Objects: This table holds entries for all the galaxies represented in the  database. Galaxies are defined by their coordinates and by name. They are created on the fly during the upload and users can specify the source with RA and DEC and optionally assign it a name. The name field also allows users to search for already defined objects. If the new object is within 2\arcsec\ of a previously defined object it is instead assigned to that.
    \item Measurements: This table holds all the results of the the automated analysis: fluxes, kinematic properties, etc. It is separated from the uploaded data so that -- in the eventuality that a major error is discovered in the analysis -- this table can be safely cleared and reconstructed without endangering the uploaded data.
\end{enumerate}

The first two tables are connected via a many-to-one relation meaning that one object may have multiple associated spectra but not the reverse. We designed this structure to accommodate the possibility that any given galaxy may have multiple observations with different instruments or settings. Each observation is then connected to one set of measurements using an estimated redshift (see Section \ref{sec:analysis}), and one set using an independently obtained systemic redshift if it is supplied by the user. 
Note that this implies that we explicitly allow for several \lya\ spectra to be uploaded for a single galaxy. This is useful if, e.g., an object has been observed with several instruments or different extraction routines are used. It is hence the responsibility of the user to compile a statistically relevant sample (for the individual usage case) from the LASD.

The web interface allows for upload of single objects as well as  multiple observations at once in tarball-format and also allows for the observations and measurements to be downloaded. We have added the possibility for uploaders to mark spectra as non-downloadable if they wish to keep the original source files proprietary but the LASD automated measurements will nevertheless be included in the downloaded measurements summary. Apart from required parameters such as a redshift estimate and a reference for the spectrum, we also allow the user to provide some additional optional information such as star formation rates and a gravitionational lensing magnification estimate.

The full pipeline that a spectrum goes through is visually represented in Figure\,\ref{fig:lasd pipeline} and each step is described in more detail in the following sections.

\begin{table*}
  \centering
  \caption{Description of spectral analysis quantities.}
\begin{tabular}{lp{9cm}l}
\toprule
                         Name &                                                                          Description &         Units \\
\midrule
             \texttt{Dx\_max} &                                Peak separation between maximum luminosity densities &          km/s \\
            \texttt{Dx\_mean} &                                 Peak separation between first moments of both sides &          km/s \\
                  \texttt{EW} &                                                            Restframe quivalent width of line &             \AA \\
           \texttt{FWHM\_max} &                                          Full-width at half maximum of highest peak &          km/s \\
           \texttt{FWHM\_neg} &                                             Full-width at half maximum of blue side &          km/s \\
           \texttt{FWHM\_pos} &                                              Full-width at half maximum of red side &          km/s \\
             \texttt{F\_cont} &                                                                  Level of continuum &  erg/s/(km/s) \\
               \texttt{F\_lc} &                                                   Luminosity density at line center &  erg/s/(km/s) \\
              \texttt{F\_max} &                                                  Luminosity density of highest peak &  erg/s/(km/s) \\
         \texttt{F\_neg\_max} &                                     Luminosity density of highest peak on blue side &  erg/s/(km/s) \\
         \texttt{F\_pos\_max} &                                      Luminosity density of highest peak on red side &  erg/s/(km/s) \\
           \texttt{F\_valley} &                                         Luminosity density of minimum between peaks &  erg/s/(km/s) \\
              \texttt{L\_neg} &                                                             Luminosity of blue side &         erg/s \\
              \texttt{L\_pos} &                                                              Luminosity of red side &         erg/s \\
              \texttt{L\_tot} &                                                                    Total luminosity &         erg/s \\
      \texttt{R\_F\_cut\_neg} &       Ratio of maximum luminosity density and peak detection threshold on blue side &               \\
      \texttt{R\_F\_cut\_pos} &       Ratio of maximum luminosity density and peak detection threshold on blue side &               \\
       \texttt{R\_F\_lc\_max} &                  Ratio of luminosity density at line center and maximum peak height &               \\
      \texttt{R\_F\_pos\_neg} &                                    Ratio of luminosity density at red and blue pea. &               \\
   \texttt{R\_F\_valley\_max} &  Ratio of luminosity density in the `valley` between the peaks and the maximum peak &               \\
      \texttt{R\_L\_cut\_neg} &                           Ratio of blueward luminosity and peak detection threshold &               \\
      \texttt{R\_L\_cut\_pos} &                            Ratio of redward luminosity and peak detection threshold &               \\
      \texttt{R\_L\_pos\_neg} &                                           Ratio of redward over blueward luminosity &               \\
              \texttt{W\_std} &                                      Square-root of second moment of whole spectrum &          km/s \\
         \texttt{W\_neg\_std} &                         Blue peak width as measured by square-root of second moment &          km/s \\
         \texttt{W\_pos\_std} &                          Red peak width as measured by square-root of second moment &          km/s \\
 \texttt{neg\_peak\_fraction} &                                          Fraction of times a blue peak was detected &               \\
 \texttt{pos\_peak\_fraction} &                                           Fraction of times a red peak was detected &               \\
                \texttt{skew} &                          Pearson's moment coefficient of skewness of whole spectrum &         \\
           \texttt{skew\_neg} &                               Pearson's moment coefficient of skewness of blue side &                        \\
           \texttt{skew\_pos} &                                Pearson's moment coefficient of skewness of red side &                        \\
              \texttt{x\_max} &                      Highest peak position determined by maximum luminosity density &          km/s \\
             \texttt{x\_mean} &                                                            First moment of spectrum &          km/s \\
         \texttt{x\_neg\_max} &                 Peak position determined by maximum luminosity density on blue side &          km/s \\
        \texttt{x\_neg\_mean} &                              Peak position determined by weighted mean on blue side &          km/s \\
         \texttt{x\_pos\_max} &                  Peak position determined by maximum luminosity density on red side &          km/s \\
        \texttt{x\_pos\_mean} &                               Peak position determined by weighted mean on red side &          km/s \\
           \texttt{x\_valley} &                                              Position of `valley` between the peaks &          km/s \\
                   \texttt{z} &                                                         Systemic redshift of source &               \\
\bottomrule
\end{tabular}

  \label{tab:description}
\end{table*}

\subsection{Initial filtering}
Before a spectrum is added to the database some tests are run to make sure that the spectrum is suitable and that we will be able to make robust measurements. Note that no manual inspection of the spectra is performed, and the integrity of the \lya\ spectra, and their identification as actual \lya\ emission lines, is left to the user. First the spectrum is converted to standard units of \AA\ for wavelength and erg/s/cm$^2$/\AA\ for flux density and the lensing magnification factor is divided out, if these parameters are given by the user. Then the following filtering steps are applied:
\begin{enumerate}
	\item First the spectral file is checked for basic consistency, such as a monotonic wavelength solution, and that no negative errors are present since negative errors indicate that the data is not trustworthy.
	\item Next, the redshift given during upload is used to isolate a 2000 km/s broad region centered on the \lya{} line. In this region the error vectors are checked against a criterion for sufficient `good data': that less than 20 percent of the values that are identically 0 since such values can be indicative of problems with the detector and can cause problems for our algorithms.
	\item Then the region $\pm 2500$ km/s around \lya\ is checked to make sure that the spectrum contains a \lya\ emission line. We also check that the spectrum is not dominated by \lya\ absorption, by requiring that the error-weighted median of the edges of the $\pm 2500$ km/s window is smaller than the error-weighted median of the central $\pm500$km/s. 
	\item The last filtering step is to check that the \lya\ peak has sufficient signal-to-noise for processing to be meaningful. In order to do this we calculate the signal-to-noise ratio of the continuum subtracted (see \S\ref{sec:analysis} for details on the continuum subtraction) spectrum in a sliding 250km/s broad window across the full $\pm 2500$ km/s spectral range. We require a minimum SNR of 7 for the spectrum to be analyzed and included. 
\end{enumerate}

\subsection{Analysis}\label{sec:analysis}

The analysis for each spectrum consists of the following steps:
\begin{enumerate}
\item continuum subtraction,
\item redshift estimation,
\item computation of the spectral quantities.
\end{enumerate}

For the \textit{continuum removal}, we first take an iterative approach:  we clip the data points that are $5\sigma$ below or above the median flux level $20$ times, and the median of the remaining points is taken as the continuum estimate. Due to the presence of a peak, and the resulting skewed flux distribution, this estimate, however, is usually too large. We therefore refine this guess by masking the region around the peak\footnote{Specifically, we mask the region $[v_a-100\kms,\,v_b+100\kms]$ where $v_a$ and $v_b$ fulfill $F(v_a) < \tilde F_{\mathrm{c}}-\sigma$ and $F_{a-1} < F_a$ and inversely for $v_b$. Here, $\tilde F_{\mathrm{c}}$ and $\sigma$ are the first guess of the continuum level and the standard deviation of the flux.} and taking the median flux of the remaining spectrum (weighted by the inverse of the error).

Estimating the systemic redshift using only the \Lya\ profile is a non-trivial problem, and it, as well as its implications, has been discussed in the literature \citep{2005ApJ...629..636A,Steidel.2010,2012ApJ...750...67R,2018MNRAS.478L..60V,2019MNRAS.489.3472B}. This is naturally due to the complicated diffusion in frequency and space \Lya photons undergo.
Additional complications include, e.g., spatially varying intrinsic \Lya spectra (as probed by, e.g., \halpha) combined with non-isotropic \Lya\ escape which makes even the definition of systemic redshift not unique.

To circumvent these problems, we chose to apply a simple definition which primarily characterizes a red and a blue peak in double peaked spectra in order to measure their quantities separately (see below). To do so, we choose the systemic redshift to be at the minimum between the two peaks in a double peaked spectrum, and blueward of the peak (thus, defining it to be the red peak) in a single peaked spectrum.
This allows us to obtain, for instance, a natural red or blue peak width while at the same time recovering the redshift of \Lya\ emitting galaxies with known systemic redshift with satisfactory accuracy (cf. Fig.~\ref{fig:redshift_comparison} and below).

In detail, the \textit{redshift estimate} works by first running a peak detection algorithm:  we use the method employed in \citet{2016ApJ...826...14G} which is a modified version of a peak detection algorithm\footnote{\url{https://gist.github.com/sixtenbe/1178136}} in conjunction with a minor Gaussian smoothing with the width of 1 resolution element to reduce high frequency  noise. 
The algorithm flags a peak (a valley) if the following $N$ data points  are at least a value of $\delta=2.5$ times the error in this region smaller (greater) than the candidate, and the minimum peak width is $7$ data points. For our purpose, we executed the algorithm for $N = (4, 6, 7, ..., 15)$ with the final result being the mode of the detected number of peaks. We constrain the separation between the peaks to be larger than $ 50\, \kms$, and smaller than $ 1200\, \kms$, and valleys are required to be surrounded by two peaks.
If two peaks are detected in the spectrum, we use the valley between the peaks as the $v=0$ estimate. 

If only a single peak is detected we employ a simple iterative algorithm on the non-smoothed spectrum for finding the estimated systemic velocity. 
First we assume the highest point in the spectrum to be the red peak. We then use a $120 \kms$ wide sliding window to select the first spectral pixel that is no longer descending as line center. 
Specifically we select the pixel that is lower than the minimum of all other blueward pixels within the window plus their error. 

For both the continuum removal and the redshift estimate, we explored a variety of different algorithms and parameter combinations and found that the ones described here work well. Note that if the true systemic redshift of a spectrum is supplied at upload, we still carry out the redshift estimation and subsequent analysis.  In these cases the LASD will estimate all the spectral analysis quantities (see below) using both the measured and estimated \zsys, and stores the measurements in two tables.  This  allows for a comparison of the resulting spectral quantities, homogenization of methods, and an evaluation of the applied redshift estimation algorithm.

For each spectrum we compute a range of  \textit{spectral quantities}, summarized in Table~\ref{tab:description}. They can be grouped in five categories:
\begin{enumerate}
\item Global quantities such as the continuum level ($F_{\mathrm{c}}$), the luminosity density at line center ($F_{\mathrm{lc}}$), the total luminosity ($L_{\rm tot}$) or the equivalent width ($EW$) of the spectrum.  They are given in units of $\mathrm{erg}\,\mathrm{s}^{-1}\,(\mathrm{km}/\mathrm{s})^{-1}$, $\mathrm{erg}\,\mathrm{s}^{-1}$,  and \AA.
\item Peak positions (starting with the \texttt{x\_} prefix) and the resulting differences (\texttt{Dx\_}). We define these positions as the point of the maximum luminosity density on the red/blue side (\texttt{\_max}) as well as the first moment of the (continuum removed) flux distribution (\texttt{\_mean} suffix). They are given as velocities in \kms.  
\item Maximum luminosity densities (\texttt{F\_}) and luminosities of the blue / red side (\texttt{L\_}). If two peaks exist, we also report the luminosity density on the `valley' between the peaks (\texttt{F\_valley}). Apart from the absolute values, we also report some ratios between them (\texttt{R\_}) which are a useful for comparison. They are given in the same units as the `global quantities' above.
\item The width of the peaks for which we use the full-width at half maximum (\texttt{FWHM\_}) as well as the second moment of the continuum subtracted flux distribution (\texttt{W\_}). Again, luminosities are given in $\mathrm{erg}\,\mathrm{s}^{-1}$ and luminosity densities are given in $\mathrm{erg}\,\mathrm{s}^{-1}\,(\mathrm{km}/\mathrm{s})^{-1}$.
\item We also compute the skewness of each peak for which we use Pearson's moment coefficient of skewness, i.e., 
\begin{equation}
    \gamma_1 = \frac{\sum_i [(x_i - \bar x) / \sigma]^3 F_i }{\sum_i F_i}
\end{equation}
where the sum is taken over the red / blue side and $\bar x$ ($\sigma$) are the first (square root of the second) moment of that side. 
\end{enumerate}

\begin{figure}
      \centering
      \includegraphics[width=.9\linewidth]{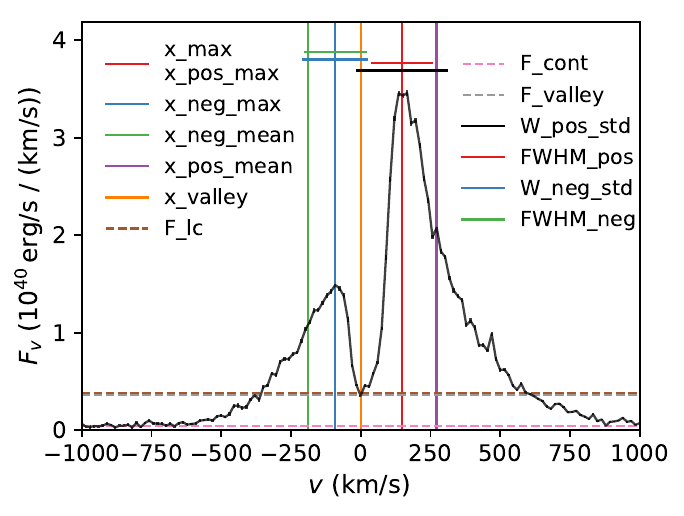}
      \caption{Visual representation of some of the measured LASD spectral quantities. Vertical lines (marked on the left with \texttt{x\_*}) show detected peaks / valleys, horizontal lines (marked on the right) show detected widths. Dashed horizontal lines show detected flux levels. 
      }
      \label{fig:spectral_quantities}
    \end{figure}

Figure~\ref{fig:spectral_quantities} shows a visual representation of the some of these measurements. We elected to use purely non-parametric properties (such as moments and weighted luminosity densities) as opposed to parametric fitting for several reasons \citep[see also discussion in][and references therein]{2020A&A...642A..55H}. The primary reason is that we require the LASD analysis pipeline to be fully automatic and ensuring the stability of non-supervised parametric model fits is non-trivial. The second reason is that the large variety of spectral profiles that are seen in \lya\ is difficult to capture in parametric models especially when model selection and tweaking needs to happen in a non-supervised fashion. Additionally this complexity leads to disagreement in what functional shapes best model the line. 

In order to quantify the uncertainty of the computed spectral quantities, we repeat the calculation $100$ times and in between `shuffle' the spectrum. 
That is, we draw a new flux in each bin from a Gaussian with mean and standard deviation being the reported flux and error, respectively. We then repeat the redshift estimation process, and if the systemic redshift (and uncertainty) is given by the user, draw a new redshift from a Gaussian defined by these values.

Ultimately, this procedure yields \textit{(i)} a redshift estimate plus uncertainty\footnote{For each measurement we report the $16$th, $50$th and $84$th percentiles as well as the value obtained from the unaltered uploaded spectrum.}, \textit{(ii)} a set of spectral quantities using this computed systemic redshift as well as their uncertainties, and, if an independent systemic redshift has been uploaded by the user, \textit{(iii)} another set of these quantities plus uncertainties.

The database is specifically designed to hold observational spectra but the same analysis of simulated \lya\ spectra will enable simple comparison between observations and simulations. For this reason we have made the analysis pipeline applied by the LASD available as an open source software package\footnote{\href{https://bitbucket.org/lya_ita/lasd_analysis}{https://bitbucket.org/lya\_ita/lasd\_analysis}}.

\section{Initial Dataset}\label{sec: Dataset}
  We initially populate the LASD with a large number of \lya\ spectra from two main archival sources, which we describe here.  We use two of the largest repositories of publicly available data, with the aim to cover both low and high redshifts with relatively homogeneous data.  At the low-$z$ end we use data obtained with the \emph{Cosmic Origins Spectrograph} (COS; \citealt{Green.2012}) onboard the \emph{Hubble Space Telescope}, obtained through the Barbara A. Mikulski Archive for Space Telescopes (MAST)\footnote{http://archive.stsci.edu/index.html}.   For high-$z$ galaxies we use publicly available data obtained with the \emph{Multi-unit Spectroscopic Explorer} (MUSE; \citealt{Bacon.2010}), mounted at Unit Telescope 4 of ESO's \emph{Very Large Telescope} (VLT), obtained through the \emph{VizieR} database\footnote{http://vizier.u-strasbg.fr/} \citep{vizier}.  
  These are also the same spectra analyzed in \citet{Hayes2020}, for which preliminary versions of the LASD software were also used.
  We stress that while these samples are large and comprise various selection functions, they are neither complete nor unbiased.  We now discuss the HST and VLT spectra in turn. 

  \subsection{HST/COS spectra at $z<0.44$}\label{sect:sample:cos}
    All the low-$z$ galaxies were pre-selected for observation based upon known characteristics, and have the advantage of having well-measured spectroscopic redshifts, usually derived from optical line emission.  The COS has targeted hundreds of galaxies with numerous General Observer (GO) and Guaranteed Time Observations (GTO)  programs, using various spectral settings.  The most common of these setting are the medium resolution gratings G130M and G160M, which span wavelengths of approximately 1150--1450~\AA\ and 1350--1750~\AA, depending upon the elected central wavelength setting (CENWAVE).  This places an upper limit on the \lya\ redshift of $\simeq 0.44$, although there is a natural bias towards lower-$z$ that results from various sample-selection and sensitivity issues.  As the Earth's upper atmosphere also glows in \lya\ (with higher surface brightness than any astrophysical source), all G130M spectra are contaminated by a geocoronal \lya\ emission feature at $\lambda = 1215.67$~\AA. We therefore place a lower limit on the recession velocity of our targets of $2500$~\kms\ in order to separate \lya\ from the geocoronal feature, although in practice the lowest redshift system included is Haro\,11 with $z=0.02$ (6000~\kms). Our sample comprises data from the following surveys, in approximately chronological order of observation: 

    \begin{itemize}

    \item{GO\,11522 and 12027 (PI: Green).  
    These galaxies stem from the COS GTO programs to study \lya\ in low-$z$ (0.02--0.06) starburst galaxies, from the Kitt Peak International Spectroscopic Survey \citep{Salzer.2001}. 
    They were primarily \halpha-selected, have star-formation rates of $\approx 0.1$ to 10~\msunyr, and 
    \lya\ is captured by the G130M grating.  They were first published in \citet{Wofford.2013}. }

    \item{GO\,11727 and 13017 (PI: Heckman). 
    These galaxies were observed in order to understand the UV properties (e.g. stellar continua and interstellar absorption lines and wind/outflows) in low-$z$ objects ($0.09< z < 0.21$) with properties analogous to those of Lyman Break Galaxies.   They were selected from the GALEX and SDSS surveys to overlap with LBGs in terms of their SFRs ($\approx 0.3$ to 60~\msunyr), UV compactness, and metallicity.  They were observed with both G130M and G160M gratings, and spectra are published in \citet{Heckman.2011,Heckman.2015}. }

    \item{GO\,12269 (PI: Scarlata). 
    This sample is the only low-$z$ study that was originally selected by  \lya-emission, which was obtained using slitless spectra from the GALEX satellite \citep{Cowie.2010,Cowie.2011}.  They were observed with COS in order to study the \lya\ emission profiles at higher spectral resolution with the G160M grating, lie at $0.19 < z < 0.34$, and have SFRs of $\approx 1-100$~\msunyr.  A stack of all these spectra is presented in Figure~8 of \citet{Songaila.2018}.}

    \item{GO\,12583 (PI: Hayes). 
    These galaxies were selected in order to study the \lya\ morphology with HST imaging, as part of the \emph{Lyman alpha Reference Sample} \citep[LARS;][]{Hayes.2014,Ostlin.2014}.  They were originally selected from SDSS and GALEX to span a range of UV luminosities comparable to LBGs. They lie at $0.029 < z < 0.18$, have SFRs of $\approx 1-100$~\msunyr, and the spectra (obtained with the G130M grating) were first published in \citet{Rivera-Thorsen.2015}. }

    \item{GO\,12928 (PI: Henry).
    These galaxies were selected from the first catalogs of starbursts known as `Green Peas' \citep{Cardamone.2009} which are particularly compact (hence `peas') and show exceptionally high equivalent width of optical [O~{\sc iii}]+\hbeta\ emission lines (giving them a green observed color at $0.18 \lesssim z \lesssim 0.44$).  They were followed up with COS to study the \lya\ profiles and outflows/winds.  Because of this selection, they occupy a narrow range in SFRs and metallicities (SFR = 5--25~\msunyr; $12+\log(\mathrm{O/H})\approx 7.9-8.1$); spectra (G160M for \lya) are published in \citet{Henry.2015}.}

    \item{GO\,13293 and 14080 (PI: Jaskot).
    The aim was to study the \lya\ emission and proxies for the neutral gas column density (as a proxy for the escape of ionizing radiation) in a sample of green pea galaxies with exceptionally ionizing stellar populations (defined by having very high [O~{\sc iii}]/[O~{\sc ii}] line ratios in the optical).  They have redshifts of $ 0.027 < z < 0.14$ which places \lya\ in both the G130M and G160M gratings, depending upon redshift and in turn program ID.  These spectra are published in \citet{Jaskot.2014} and \citet{Jaskot.2017}. }

    \item{GO\,14201 (PI: Malhotra)
    These galaxies are also a sub-set of the green peas, and were selected specifically to study the \lya\ output of galaxies as a function of various other properties.  They have SFRs of 4--40~\msunyr\ and redshifts of $0.18 < z <0.33$, which places \lya\ in the G160M grating.  Spectra are published in \citet{Yang.2017}, although note that this paper also compiles spectra from many of the programs mentioned above, including 11727, 12928, and 13293.}

    \item{GO\,13744 (PI: Thuan), 14635, and 15136 (PI: Izotov).  The first two programs were designed to study the ionizing emission from Green Pea galaxies (13744) and GPs with extreme [O~{\sc iii}]/[O~{\sc ii}] ratios (14635).  This places them at somewhat higher redshifts, $z=0.29-0.43$ and redshifts \lya\ into the G160M grating.  All these galaxies emit a substantial fraction of their Lyman continuum radiation. The final program was designed to study the \lya\ emission from similar objects (15136), but concentrated at lower-$z$ (0.03--0.07), placing \lya\ in G130M.
    These galaxies have SFRs of 15-40~\msunyr\ and spectra are published in \citet{Izotov.2016,Izotov.2018,Izotov.2020}.}

    \end{itemize}

    We are mainly concerned about the \lya\ emission from star-forming galaxies, and do not consider programs targeting active-galactic nuclei, AGN (or those where the probability of AGN inclusion is high; e.g. GO\,12533 and 13407, PI: Martin).  There are also a number of galaxies with \lya\ data from COS, but for which only low resolution spectra have been obtained with the F140L setting.  We do not consider these spectra for the initial population of the database. 

    We obtained all these data from the MAST archives, reprocessing everything homogeneously with Version 3.3.7 of the 
    calibration pipeline (\texttt{CALCOS}).  We first check the centering of the galaxies in the COS near ultraviolet acquisition images, and the central wavelength of the geocoronal emission lines in the extracted spectra for every integration, to ensure an accurate wavelength solution.  We reject a very small number of individual exposures that have anomalously short integration times or shutter failures.  We then use a custom script to combine the individual spectra for each system, conservatively rejecting \emph{all} spectral pixels with data quality (DQ) flags not equal to zero.  We examine the error spectrum for each spliced spectrum, and contrast it with the error spectrum expected from the galaxy spectrum and Poisson statistics; we then follow the method outlined in Section 3.3 of \citet{Henry.2015}  to recompute the error spectrum, which differs significantly from expectation in the cases of poorly exposed spectra.  We finally rebin the signal and error spectra to critically sample their native spectral resolution -- simply binning by a uniform factor of six spectral pixels  --  although ultimately this process is only aesthetic and should not affect the quantities derived by the LASD algorithms.  The final intrinsic resolving power ($R \equiv \lambda/\Delta\lambda$) varies between 13,000 and 19,000 depending upon grating, precise wavelength of redshifted \lya, COS lifetime position, and the size of the \lya-emitting region with respect to the COS aperture.  

  \begin{figure*}
    \centering
    \includegraphics[width=\linewidth]{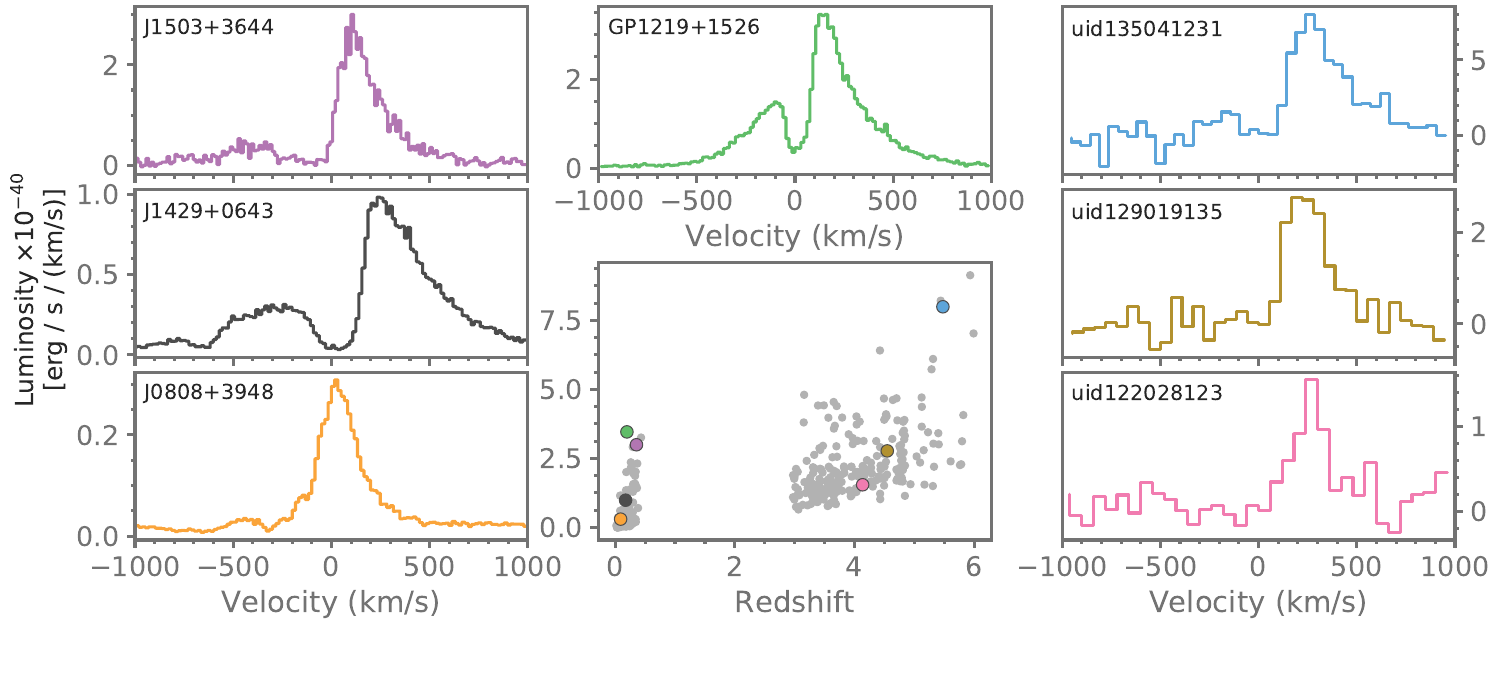}
    \caption{Central panel: Luminosity distribution of the galaxies included in the LASD as a function of redshift. Surrounding panels: Example spectra showing typical spectral profiles across a range of luminosities and redshifts. The colored points in the luminosity distribution indicate the position of the spectrum with the corresponding color. All spectra are shown in the same velocity range. All y-axes are luminosity densities in units of $10^{40}$ erg s$^{-1}$ (km/s)$^{-1}$}.
    \label{fig:spectrumshowcase}
  \end{figure*}

  \subsection{VLT/MUSE spectra at $2.9 < z< 6.6$} 

    MUSE has revolutionized high-redshift surveys for emission line galaxies since its installation at VLT.  Because of its very large number of detectors, MUSE simultaneously has a very large field-of-view ($60\times60$~\sq\arcsec), a small pixel area ($0.2\times 0.2$~\sq\arcsec), and long-baseline optical wavelength coverage ($\lambda=4800-9300$~\AA) at 1.25~\AA\ sampling.  Consequently, MUSE samples a cosmic volume of $\approx$~10,000~Mpc$^3$ for \lya-emitters in every pointing and, because of its high throughput and the 8.2~m aperture of VLT,  MUSE is a very efficient survey instrument.

    MUSE has already been used for many \lya\ emitter surveys, of various depth between short, 1-hour observing blocks and the stupendously deep field of 190 hours.  To initially populate the LASD we take the publicly distributed data from the MUSE-WIDE survey \citep{Urrutia.2019}, which comprises 44 MUSE datacubes in the CANDELS-Deep region of the GOODS-South field \citep[see also ][]{Herenz.2017mw}.  This data-release (DR1) contains 479 \lya-emitting galaxies at $z\ge 2.9$, compiled into a  catalog including emission-line selected galaxies (using the \texttt{LSDCat} software \citealt{Herenz.2017lsd}), and by the extraction of spectra from photometrically pre-selected objects \citep[e.g.][]{Guo.2013}.  We obtained all the MUSE-Wide spectra, reduced, identified and extracted by  \citet{Urrutia.2019}, from the CDS/VisieR.   For the analysis presented in this paper we further restrict ourselves to galaxies for which the lead-line is \lya\ and the integrated SNR exceeds 8. 

  \subsection{Ingestion into the LASD database}
    In principle the spectra could be uploaded to the LASD in the form in which we have hereto described.  However as the focus is on emission line profiles and kinematic signatures, we restrict our catalogs to galaxies with strong \lya\ lines/higher signal-to-noise.  Naturally this modifies the selection bias towards more luminous galaxies at a given redshift.  Specifically concerning the COS sample at low-$z$, almost none of these galaxies were selected on their \lya\ emission (only GO\,12269; PI: Scarlata) and a selection are net absorbers of \lya\ or have weak features because of high \hi\ column densities (this is mainly true for the KISSR sample of \citealt{Wofford.2013}).   For both COS and MUSE-systems, we retain only galaxies with net \lya\ emission lines, defined as line flux detected at SNR$\ge8$ in a region of $\pm 2500$~\kms\ from the systemic redshift of \lya.  This reduced the number of COS spectra from 145 to 123, as some galaxies are \lya\ absorbers.  Using the same criterion, the MUSE-Wide sample is reduced from 479 to 234 \lya-emitters, as many galaxies have SNR lower than quoted.

    The LASD can accept spectra with either systemic redshifts (i.e. measured by other emission lines) or more approximate redshifts estimated from the  \lya\ line (see Section~\ref{sec:analysis}).  For the COS samples we upload the spectra with known $z_\mathrm{sys}$, usually based upon nebular lines in the optical, where we compiled the redshifts from the papers listed in Section~\ref{sect:sample:cos}.   For COS-observed low-$z$ galaxies with SDSS spectra, we re-measure $z_\mathrm{sys}$ using 20 of the strongest optical emission lines; for the remainder we refer to measurements presented in the papers listed in Section~\ref{sec: Dataset}.  For the MUSE-Wide sample the we take the redshift estimates from \citet{Urrutia.2019}.

\section{Validations and example science cases}\label{sec: Results}

  Once we had uploaded the above datasets to the database we downloaded the resulting measurements and in this section we demonstrate some results that can be derived directly from this dataset. 
  In figure \ref{fig:spectrumshowcase} we show the distribution of luminosities of the uploaded \lya\ emitters together with some representative spectra from both the COS and MUSE samples. It is directly evident from this figure that there are a large variety of \lya\ spectral profiles in the database, ranging from double peaks to P-Cygni type profiles to single peak profiles. Single peak profiles are relatively more frequent in the high redshift samples which could be due to resolution effects. However, it could also be due to blue peaks being preferentially absorbed in the increasingly neutral IGM at high redshifts \citep[e.g.][]{Hayes2020}.

  \subsection{Redshift detection}
    \begin{figure}
      \centering
      \includegraphics[width=.9\linewidth]{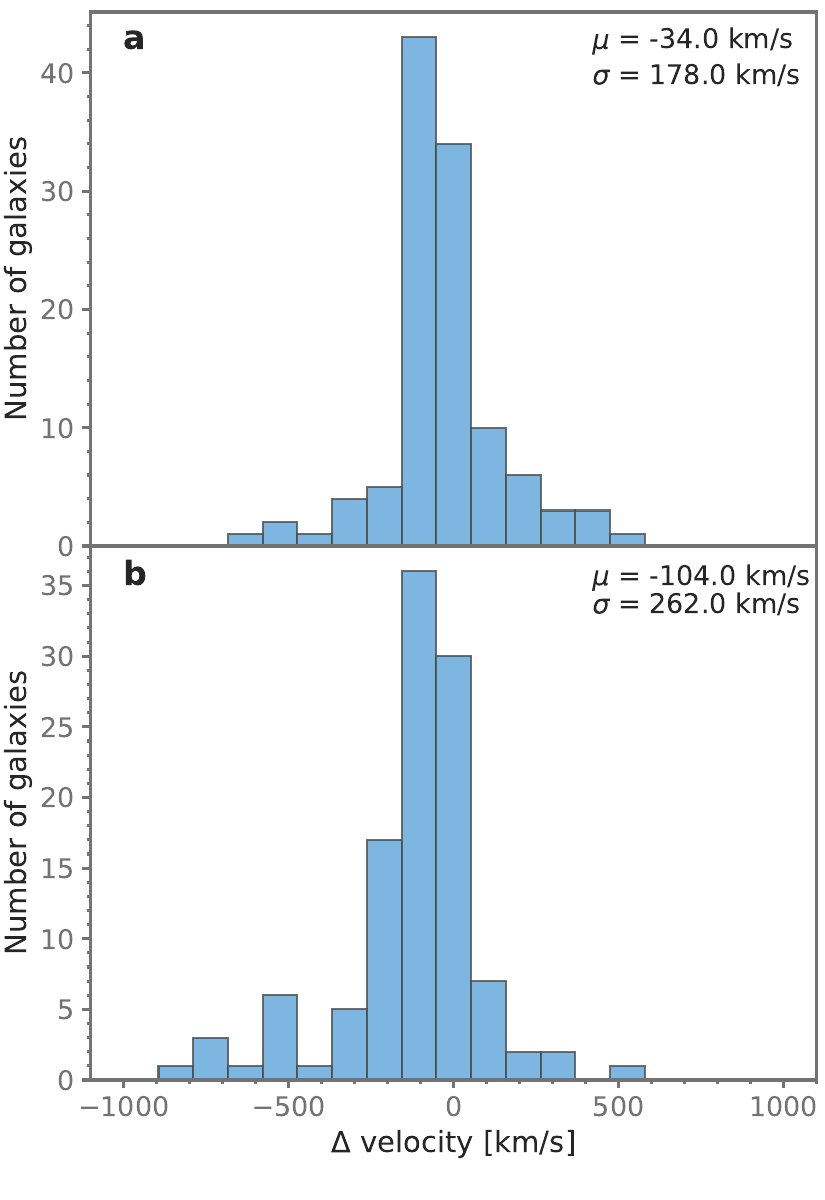}
      \caption{Panel\,\textbf{a}: Distribution of the difference of the estimated redshift and the true redshift for high resolution (R$\sim$ 17000) spectra. Panel\,\textbf{b}: as Panel\,\textbf{a} but at approximate MUSE resolution. $\mu$ indicates the mean of the distribution and $\sigma$ is the standard deviation.
      }
      \label{fig:redshift_comparison}
    \end{figure}
    One of the most crucial processes that happens in the LASD processing pipeline is redshift determination, since many high redshift galaxies lack independently determined redshifts. In our initial dataset this applies to all MUSE Wide galaxies. In order to check the accuracy of the automated redshift detection algorithm we compared the estimated redshift to the true systemic redshifts for COS sample where the redshifts are precisely and independently known from optical spectroscopy. The difference between the estimated and the true redshifts are shown in panel\,a of Figure~\ref{fig:redshift_comparison}. The differences show a relatively narrow distribution of values with a median value of -59 km/s and 25th (75th) percentile at --137(37)km/s. This indicates a slight shift towards detecting lower redshifts than true which is expected based on how our algorithm operates. Overall, however, the distribution shows no strong indications of any major systematic bias at COS resolutions.

    However, we must also take into account the fact that the redshift determination algorithm is sensitive to the spectral resolution of the spectrograph.  We cannot use the actual MUSE spectra to estimate the size of this effect since they do not have independent redshift determination. We therefore create artificial low resolution spectra by convolving the COS spectra with a kernel corresponding to R $\sim 4500$ and rebin the spectrum to the Nyquist sampling for this resolution. This kernel combined with the effective resolution of COS for the extended \lya\ emission of these low-z galaxies corresponds roughly to the spectral resolution of MUSE \citep{Hayes2020}. The resulting convolved spectra were then run through the redshift detection algorithm again, and difference between the LASD-estimated and true redshifts are shown in panel\,b of Figure~\ref{fig:redshift_comparison}.

    Panel\,b shows that at the lower resolution the distribution of differences is no longer entirely symmetric but shows a skew and a small systematic offset on the negative side. This means that for resolutions below R $\sim 5000$ we are in general finding redshifts that are slightly too low. This is what is expected since for all single peak profiles the algorithm detects the blue edge of the \lya\ line which is shifted towards the blue as the profile is broadened at lower resolution. The distribution is also somewhat broadened compared to the high resolution case which is most likely due to the impact of the large variety of spectral profiles responding differently to the spectral resolution decrease. For instance, double peak profiles that blend together and become unresolved at the lower spectral resolution will cause the left edge of the profile to move considerably blueward compared to the original valley position. The prevalence of this effect will strongly depend on the properties of the input spectrum, such as the intrinsic peak separation, and the resolution of the spectrograph. Simple testing on our high resolution sample shows that significant loss of blue peaks starts to occur below spectral resolutions of $\sim4000$ which is in agreement with the results of \citet{Verhamme2015}, see Figure \ref{fig:DB_Peaks} in Appendix \ref{app:DB}.

  \subsection{Distribution of \lya\ properties}
  In this section we present some of the distributions of \lya\ properties in our initial sample, as well as some of the correlations present in our homogeneously measured dataset. This is not an exhaustive examination of all the correlations present in the dataset, and we encourage the reader to download the data and do further explorations.
   
    \begin{figure}
      \centering
      \includegraphics[width=.9\linewidth]{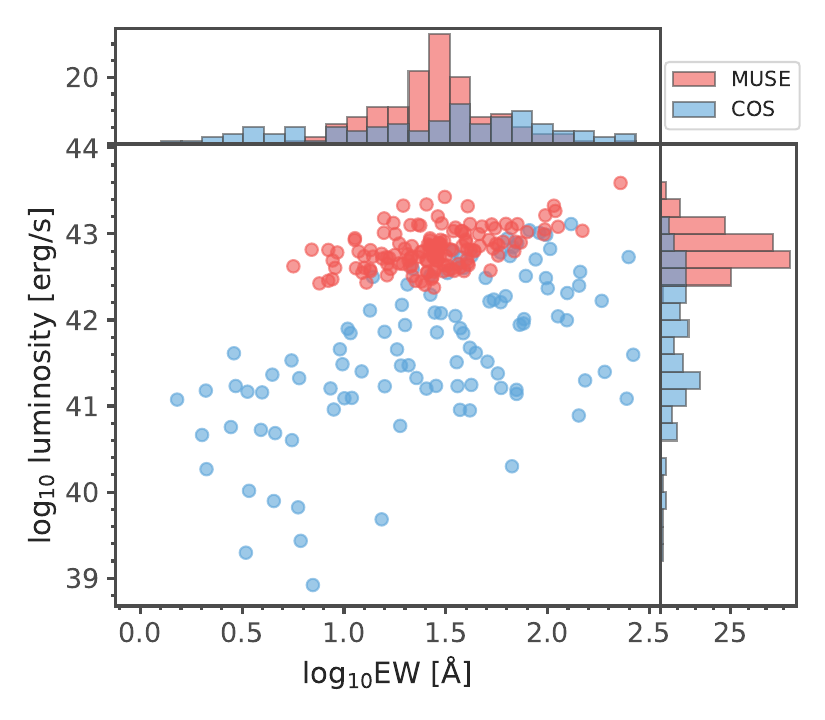}
      \caption{\lya\ luminosity and EW correlation and the distributions of each of these parameters for the low and high redshift galaxies in the LASD.
      }
      \label{fig:LvEW}
    \end{figure}
  
  Figure~\ref{fig:LvEW} shows the distributions of  total \lya\ luminosity versus the equivalent widths. We first note that the EW and the luminosity do correlate for both the low and high redshift galaxy samples. It is also clear that the low-z COS sample of galaxies samples a wider range in luminosity and EW. This is expected since the selection functions for the COS galaxies are much more diverse than the \lya\ selection of MUSE. It also seems that the slope of the Luminosity -- EW relation is shallower for the high redshift galaxies, which is likely because of the very different selection functions of the low- and high-$z$ datasets.

\begin{figure}
      \centering
      \includegraphics[width=.9\linewidth]{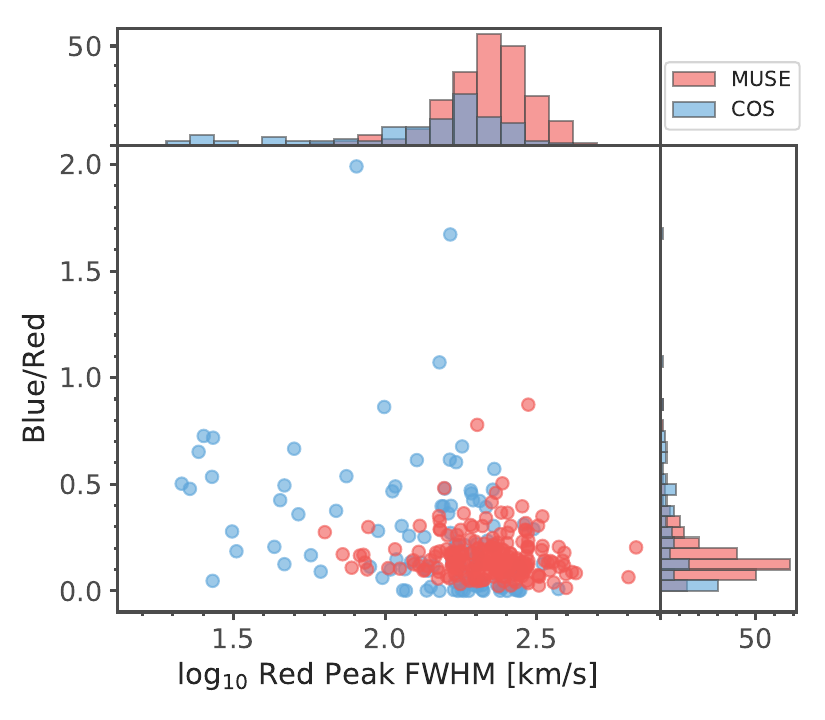}
      \caption{Ratio of the luminosity blueward of line center and the luminosity redward of line center, versus the full-width-half-maximum of the red peak. The distributions of each of these parameters for the low and high redshift galaxies in the LASD are also shown.}
      \label{fig:BRvFWHM}
    \end{figure}

  Another illustrative example of the available data is shown in Figure~\ref{fig:BRvFWHM} which shows the ratio of the luminosities blueward and redward of line center compared to the width of the red \lya\ peak. The figure again illustrates the differences between the two samples with the blue COS  galaxies showing a much larger spread, particularly of the FWHM compared to the MUSE sample. This is partly expected since the spectral resolution, R, of MUSE is much lower than that of COS, causing the line to be broadened. However the observed FWHMs range from 100 km/s to $\sim$500 km/s, which is much broader than the instrumental resolution, which is approximately $\sim$150 km/s. The most probable cause of this difference is that the COS sample contains galaxies that are significantly less luminous than those in the MUSE sample and therefore are likely to have smaller intrinsic velocity dispersions. This is also corroborated by the data which shows that the MUSE galaxies do have comparable FWHM to COS galaxies of similar luminosities.

\subsection{Limitations}
While doing homogeneous measurements for a large set of galaxies provides opportunities for unique insights into the properties of \lya\ radiation there are some limitations that are good to keep in mind when interpreting measurements and correlations. The first of these is the difficulty of accurately determining redshifts from the \lya\ spectral line. While we demonstrated that the redshift detection is robust and show no major systematic deviations across the whole sample there are still some uncertainties for a single given galaxy and this uncertainty will propagate into some of the measured quantities, such as the peak positions. The automatic redshift detection will also likely cause the fraction of luminosity on the blue side of \lya\ to be systematically underestimated. 

There is also an additional bias originating from spectral resolution effects which impact not only the redshift detection but also many of our measured quantities, such as second moments and FWHMs, directly.

We finally note that observations may be obtained with any kind of spectrograph (slits, fibers, integral field, etc).  As apertures can vary in size, and so can the atmospheric seeing, slit losses will differ from observation to observation, especially for the extended \lya\ line.  Lensed galaxies could be even more affected.  Given the number of possible choices to be made, we do not record information pertaining to aperture definition, but caution the community that aperture effects will be at play and affect the photometry at an uncertain level.

\section{Outlook}
The usage of \Lya\ in astronomy has transitioned from purely theoretical to heavily data driven. New instruments at large telescopes such as MUSE, Keck Cosmic Web Imager (KCWI; \citealp{KCWI2012,KCWI2018}), and XSHOOTER \citep{XSHOOTER2011} increased the number of observed \Lya spectra by orders of magnitudes in recent years.
Also on the theoretical side there is steady progress with new analytic solutions \citep{2006ApJ...649...14D,2016ApJ...833L..26G} and radiative transfer codes \citep{2017MNRAS.464.2963S,2020A&A...635A.154M} available to the community. 
With this progress it becomes increasingly important that the individual pieces of knowledge become better connected, i.e., that new data acquired is compared to existing one, and that theory is compared to data. 

A major hurdle to overcome is the availability of \Lya spectra. While some telescopes do have their dedicated archives, the reduced spectra are not easily accessible. Furthermore, over the years different definitions of the same quantities developed which complicate comparisons.

In this work, we have presented the Lyman Alpha Spectral Database (LASD). The database consists of a analysis software and a web portal which allows for the access of homogeneously measured \lya\ line quantities, and \Lya spectra -- as well as the upload of new spectra which will then be automatically be analysed. 
The database was designed to increase the access to comparison samples for both observational and theoretical work and to facilitate the sharing of data across research groups. 
We have populated the database with a sample of 332 archival spectra which we also present in this paper.

The LASD is intended to be a tool for the \lya\ community to use and in order for it as useful as possible we encourage the reader both to upload new spectra and to explore the LASD dataset. We highlight that when a user uploads a spectrum they have the choice to share the full spectral data, or simply the LASD measured quantities. Furthermore, we encourage the users to cite the original observational paper when using the LASD and we provide a BibTeX file containing these references for convenience. 

Given acceptance by the community, we plan to expand the LASD to feature more measurements, more built-in data exploration tools, improved links to auxiliary data, broader upload file specifications, and other improvements suggested by the users. Input from the community is both welcome and encouraged.  

\label{sec:conclusions}

\section*{Acknowledgements}
We thank Jorryt Matthee for useful discussions and testing preliminary versions of the LASD. We also thank the referee for useful comments that helped improve both the database and this manuscript.
This research has made use of NASA's Astrophysics Data System, matplotlib \citep{Hunter:2007}, SciPy \citep{Virtanen_2020}, the IPython package \citep{PER-GRA:2007}, Astropy \citep{2018AJ....156..123A,2013A&A...558A..33A}, NumPy \citep{van2011numpy,Harris2020}, pandas \citep{McKinney_2010, McKinney_2011}, and other open software.
MG was supported by NASA through the NASA Hubble Fellowship grant HST-HF2-51409 and acknowledges support from HST grants HST-GO-15643.017-A, HST-AR-15039.003-A, and XSEDE grant TG-AST180036.  M.H. acknowledges the support of the Swedish Research Council, Vetenskapsr{\aa}det, and is Fellow
of the Knut and Alice Wallenberg Foundation.



\appendix
\section{Double Peak Recovery fraction}\label{app:DB}
    \begin{figure}
      \centering
      \includegraphics[width=.6\linewidth]{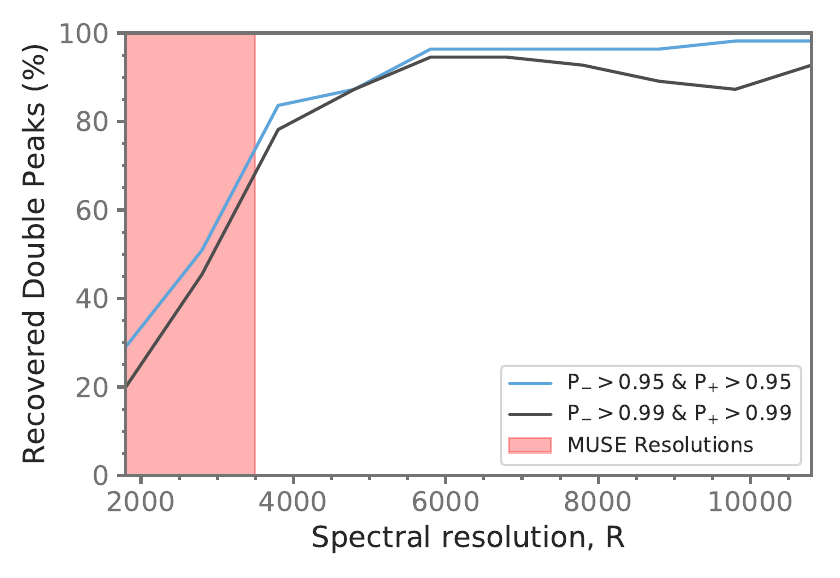}
      \caption{Illustration of the LASD algorithms ability to characterize a spectrum as double peaked as a function of the spectral resolution. The blue line shows the spectra fraction that has a blue peak and red peak detection in more than 95\% of Monte Carlo iterations and the black line shows fraction with detections in 99\% of iterations}
      \label{fig:DB_Peaks}
    \end{figure}
As a part of the analysis done to characterize the sensitivity of the redshift detection algorithm to spectral resolution, we also tested what spectral resolutions are required for double peaks to be properly classified by the LASD. The methodology was as follows: We select a sample of spectra that were classified as double peaked at COS resolutions and visually confirm these samples. Then the spectra are degraded to lower spectral resolutions using the methodology described in the main manuscript and the detection algorithm is run again. What is noticeable in Fig.~\ref{fig:DB_Peaks} is that the algorithm is largely unaffected by resolution effects until R$\sim4000$ after which the detection fraction drops precipitously. This is consistent with the results presented in \citet{Verhamme2015}.
\bibliographystyle{aasjournal}
\bibliography{refs}

\label{lastpage}

\end{document}